\documentclass[journal]{IEEEtran}
\usepackage{amsmath,amsfonts}
\usepackage{algorithmic}
\usepackage{algorithm}
\usepackage{array}
\usepackage{tikz}
\usepackage{tabularx}
\usepackage{subfigure}
\usepackage{setspace}
\newcommand*\emptycirc[1][1ex]{\tikz\draw (0,0) circle (#1);} 
\newcommand*\halfcirc[1][1ex]{%
	\begin{tikzpicture}
	\draw[fill] (0,0)-- (90:#1) arc (90:270:#1) -- cycle ;
	\draw (0,0) circle (#1);
	\end{tikzpicture}}
\newcommand*\fullcirc[1][1ex]{\tikz\fill (0,0) circle (#1);} 

\usepackage{textcomp}
\usepackage{stfloats}
\usepackage{url}
\usepackage{verbatim}
\usepackage{graphicx}
\usepackage{cite}
\usepackage{framed}
\usepackage{multirow} 
\usepackage{booktabs}
\usepackage{makecell}
\begin{document}


\title{Double Whammy: Stealthy Data Manipulation aided Reconstruction Attack on Graph Federated Learning}




\author{Jinyin Chen,
        Minying Ma,
        Haibin Zheng,
        Qi Xuan

	\thanks{J. Chen, H. Zheng  and Q. Xuan are with the Institute of Cyberspace Security, the College of Information Engineering, Zhejiang University of Technology, Hangzhou, 310023, China. (e-mail: chenjinyin@zjut.edu.cn, zhenghaibin@zjut.edu.cn,xuanq@zjut.edu.cn)}
	\thanks{M. Ma is with the College of Information Engineering, Zhejiang University of Technology, Hangzhou 310023, China. 
    E-mail:
    mmy021456@163.com.}
 	\thanks{This research was supported by Zhejiang Provincial Natural Science Foundation (No. LDQ23F020001), National Natural Science Foundation of China (No. 62072406), the Zhejiang Province Key R\&D Science and Technology Plan Project (No. 2022C01018)} 
\thanks{Corresponding author: Jinyin Chen.}
}

\maketitle

\begin{abstract}
Graph federated learning (GFL) is one of the effective distributed learning paradigms for training graph neural network (GNN) on isolated graph data. It perfectly addresses the issue that GNN requires a large amount of labeled graph without original data sharing. Unfortunately, recent research has constructed successful graph reconstruction attack (GRA) on GFL. 
But these attacks are still challenged in aspects of effectiveness and stealth.
To address the issues, we propose the first \underline{D}ata \underline{Man}ipulation aided \underline{Rec}onstruction attack on GFL, dubbed as \emph{DMan4Rec}. The malicious client is born to manipulate its locally collected data to enhance graph stealing privacy from benign ones, so as to construct double whammy on GFL.
It differs from previous work in three terms: (\romannumeral1)~\emph{effectiveness} - to fully utilize the sparsity and feature smoothness of the graph, novel penalty terms are designed adaptive to diverse similarity functions for connected and unconnected node pairs, as well as incorporation label smoothing on top of the original cross-entropy loss. (\romannumeral2)~\emph{scalability} - DMan4Rec is capable of both white-box and black-box attacks via training a supervised model to infer the posterior probabilities obtained from limited queries. (\romannumeral3)~\emph{stealthiness} - by manipulating the malicious client's node features, it can maintain the overall graph structure's invariance and conceal the attack. 
Comprehensive experiments on four real datasets and three GNN models demonstrate that DMan4Rec achieves the state-of-the-art (SOTA) attack performance, e.g., the attack AUC and precision improved by 9.2\% and 10.5\% respectively compared with the SOTA baselines. Particularly, DMan4Rec achieves an AUC score and a precision score of up to 99.59\% and 99.56\%, respectively in black-box setting. Nevertheless, the complete overlap of the distribution graphs supports the stealthiness of the attack. Besides, DMan4Rec still beats the defensive GFL, which alarms a new threat to GFL.
 
\end{abstract}


\section{Introduction}
During the last decade, graph neural network (GNN) has dominated the tasks for graph analysis, leading to its widespread application in various fields, such as recommendation systems
~\cite{DBLP:journals/corr/abs-2102-04925}, drug discovery~\cite{DBLP:conf/ijcai/WangL0Q020}, and etc~\cite{DBLP:conf/esws/SchlichtkrullKB18, DBLP:conf/iclr/ChenMX18, Qiu2018PrivacypreservingWC}.
Recently, due to privacy concerns, regulatory restrictions, and commercial competition, practical applications require the decentralization of graph data, and graph federated learning (GFL)~\cite{DBLP:journals/corr/abs-2105-11099, DBLP:journals/corr/abs-2202-07256, fu2022federated} is coming up just in time, which is perfectly designed to train the global GNN on a large amount of labeled graph without raw data sharing. To cope with different data distribution applications~\cite{Zhang2021FederatedGL}, GFL is roughly categorized into graph horizontal federated learning (GHFL)~\cite{DBLP:journals/corr/abs-2104-04141, xie2021federated, DBLP:journals/corr/abs-2102-04925, liu2022federated} and graph vertical federated learning (GVFL)~\cite{DBLP:journals/corr/abs-2005-11903, DBLP:journals/corr/abs-2106-11593, DBLP:conf/ijcai/0001ZZWLWWLWZ22, DBLP:conf/icdm/Wang0PL022}. 
Since GHFL captures the mainstreaming attention in practice~\cite{wang2024horizontal}, thus we focus on GHFL and refer to it as GFL for short in this paper.

\begin{figure}[htbp!]
	\centering
	\includegraphics[width=1\linewidth]{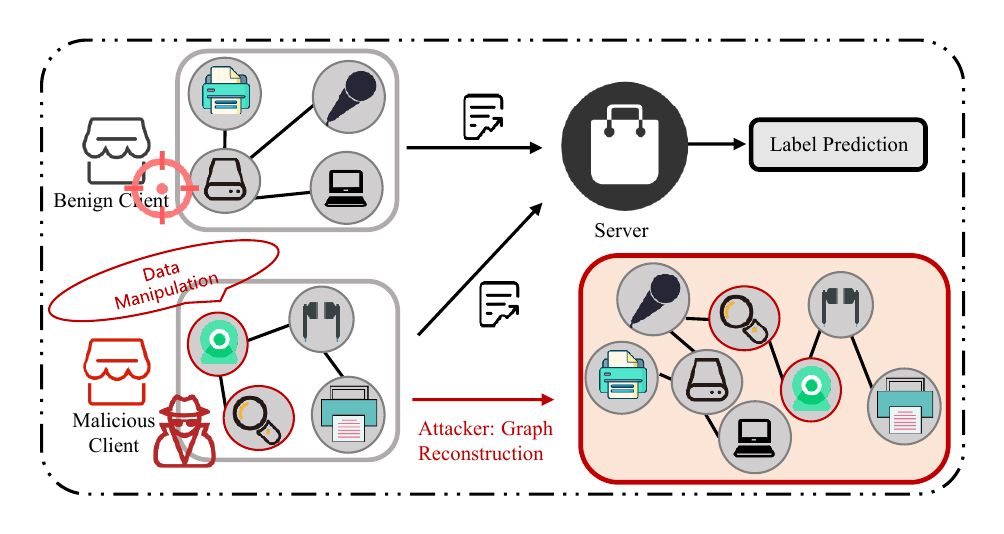}\\
	\caption{An example of the graph reconstruction attack against the Amazon shopping website based on data manipulation. }
	\label{fig1}
\end{figure}

Although GFL is proposed to train a global model with isolated data, unfortunately recent research~\cite{10.1145/3543507.3583359, wu2021fedgnn} has launched successful inference attack on GFL, which has revealed its vulnerability towards privacy leakage. Graph reconstruction attack (GRA) is one of the inference attacks that can compromise the intellectual property, confidentiality, or privacy of graphs. Therefore, we specifically focus on GRA against GFL~\cite{ijcai2021p516}. In the case of Amazon's shopping site, different stores on this platform sell various products, each maintaining its own graph networks, making it suitable for GFL scenarios. In this context, competing stores may sell similar types of products. Stores infer the complete graph network and obtain more product information through a federated training process as shown in Fig.~\ref{fig1}. In this way, the malicious store can adjust its marketing strategy to enhance the exposure and recommendation level of their products in the recommendation system. More importantly, taking advantage of the inherent ease of manipulating local data in GFL, the malicious store can conveniently access and modify its locally collected data~\cite{9148790} without sharing it with other clients or the server. This enables a greater ability to steal more private information from benign clients.

Currently, numerous studies have focused on GRA within graph-related fields, but these attacks~\cite{ijcai2021p516, 277160, 10386501} fail to steal privacy in GFL scenarios since they are launched in assumption of centralized training. In this context, we summarized the main challenges of GRA on GFL, i.e., (\romannumeral1)~\emph{effectiveness:} due to the sparsity of graphs, joint training across clients exacerbates this vulnerability,  thereby preventing the attack from being fully effective; (\romannumeral2)~\emph{scalability:} in the GFL scenario, the malicious client lacks direct access to the server model, as well as they can only query the node posterior probability, resulting in difficulty to construct black-box attack; (\romannumeral3)~\emph{stealthiness:} the existing methods of manipulated data, such as malicious node injection~\cite{fang2024gani}, rewiring~\cite{chen2020mga}, and link modification~\cite{bojchevski2019adversarial, xu2019topology, ma2019attacking, zang2020graph} are purposely designed for backdoor injection in general, that is significantly increase the risk of being detected by defensive mechanism.

In order to solve the aforementioned issues and uncover possible security vulnerabilities for GFL scenarios, we introduce a novel data manipulation aided reconstruction attack (DMan4Rec) for the first time. Specifically, to tackle the challenge of effectiveness and ensure that the manipulated graph enhances the similarity of GNN outputs for connected nodes while promoting dissimilarity for unconnected nodes, we design adaptive penalties for both connected and unconnected node pairs by employing various similarity functions. In addition, to improve the generalization performance of the model, we incorporate label smoothing into the original cross-entropy loss to fully leverage feature smoothness of the graph. In order to address the scalability challenge, a supervised model is trained using only the posterior probability obtained by querying the server model and the shadow dataset obtained using malicious nodes, thus converting the attack into a supervised classification problem. Since existing methods~\cite{fang2024gani, chen2020mga, bojchevski2019adversarial, xu2019topology, ma2019attacking, zang2020graph} for manipulating data are specifically designed for backdoor injection, they significantly increase the risk of detection by defensive mechanisms. Therefore, we strategically manipulate node features to contaminate the graph while carefully preserving its overall structure, addressing the challenge of stealthiness.

In summary, the main contributions are outlined as follows:
\begin{itemize}
\item 
\emph{Attack scenario.} Since the natural characteristics of distributed data in GFL, malicious client is easy to manipulate local data, we propose the first stealthy data manipulation aided graph reconstruction attack on GFL, dubbed as DMan4Rec. It outperforms previous work in aspects of effectiveness, scalability and stealthiness.

\item
\emph{Attack framework.}  A novel penalty terms is proposed to adaptive to diverse similarity function, as well as incorporation label smoothing strategy for attack effectiveness. DMan4Rec is scalable of both white-box and black-box attacks by achieving the state-of-the-art (SOTA) attack performance. The malicous client manipulates limited  node features to maintain the overall graph structure’s invariance to promise the stealthiness.

\item 
\emph{Attack performance.} Extensive experiments are conducted on 4 general graph datasets, 3 GNN models and 5 baselines, and the results testify that DMan4Rec improves AUC and precision by 9.2\% and 10.5\% respectively compared with the SOTA baselines, without affecting classification performance. Particularly, in balck-box scenario, DMan4Rec can achieve an AUC score and a precision score of up to 99.72\% and 99.96\%, respectively, under the distribution graphs before and after data manipulation completely overlap, ensuring the concealment of the attack. Additionally, DMan4Rec can also beat the defensive GFLs.

\end{itemize}

The rest of the paper is organized as follows. Related works are introduced in Section \uppercase\expandafter{\romannumeral2}, while the proposed method is detailed in Section \uppercase\expandafter{\romannumeral3}. Experimental results and discussion are showed in Section \uppercase\expandafter{\romannumeral4}. At last, we conclude our work and point future work.

\section{Related Work}
In this section, we will briefly review  existing works on several aspects, i.e., graph federated learning (GFL), graph reconstruction attack (GRA), and poisoning attacks against GFL scenarios.

\subsection{Graph Federated Learning (GFL)}
GFL~\cite{DBLP:journals/corr/abs-2105-11099} is a novel distributed learning paradigm that combines the ideas of federated learning and GNNs to support new opportunities for privacy preserving distributed graph learning. The existing research on GFL primarily focus on horizontal~\cite{DBLP:journals/corr/abs-2104-04141, xie2021federated, DBLP:journals/corr/abs-2102-04925, liu2022federated} and vertical scenarios~\cite{DBLP:journals/corr/abs-2005-11903, DBLP:journals/corr/abs-2106-11593, DBLP:conf/ijcai/0001ZZWLWWLWZ22, DBLP:conf/icdm/Wang0PL022}.

\textbf{Graph horizontal federated learning (GHFL).}
GHFL refers to an federated learnign (FL) setting in which participants share the same feature and label space but operate within different node ID spaces. Wang \emph{et al.} \cite{DBLP:journals/corr/abs-2104-04141} proposed a computationally efficient method for searching graph convolutional network (GCN) architectures within the FL framework, enabling the identification of superior GCN models in a shorter time. Regarding various sets of graphs, Xie \emph{et al.} \cite{xie2021federated} proposed a graph clustered federated learning framework that dynamically identifies clusters of local systems based on the gradients of GNN. However, this work primarily emphasizes the effectiveness of FL in this setting without thoroughly examining other critical issues, such as data privacy. More concentrated in one direction, federated frameworks~\cite{DBLP:journals/corr/abs-2102-04925},\cite{liu2022federated} were proposed for privacy preserving GNN based recommendations.

\textbf{Graph vertical federated learning (GVFL).}
GVFL is an FL setting in which multiple parties share the same node ID space but have different feature and label space. Zhou~\emph{et al.}~\cite{DBLP:journals/corr/abs-2005-11903} put forward the first vertical learning paradigm for privacy preserving GNN models for node classification by partitioning the graph into two segments for different clients. Additionally, secure multi-party computation is employed to ensure data privacy and efficiency. Ni~\emph{et al.}~\cite{DBLP:journals/corr/abs-2106-11593} proposed a vertical federated learning framework named FedVGCN, and to ensure privacy under this framework, it adopted additively homomorphic encryption (HE). Chen~\emph{et al.}~\cite{DBLP:conf/ijcai/0001ZZWLWWLWZ22} further proposed some novel combination strategies for the server to combine local node embeddings from graph data holders in FL so as to improve the effectiveness. As for the semi-supervised node classification task, Wang~\emph{et al.}~\cite{DBLP:conf/icdm/Wang0PL022} incorporated model-agnostic meta-learning into GFL to handle non-IID graph data, while preserving the model’s generalizability. 

Most proposed GFL methods either concentrate on enhancing model efficiency, or prioritize privacy protection at each step. However, most of them overlook the risk of privacy disclosure. Consequently, we focus on GHFL due to its increasing prevalence in practice~\cite{wang2024horizontal}, and refer to it simply as GFL.

\subsection{Graph Reconstruction Attack (GRA)}
\setlength{\parskip}{0.01cm plus2mm minus2mm}
GRA refers the attacks aiming at reconstructing a graph that possesses similar structural properties, such as graph structure, degree distribution, and local clustering coefficient, to those of a target graph. Depending on whether these methods manipulate data, they can be categorized into two types, i.e., GRAs with malicious data and GRAs without malicious data. For GRAs without malicious data, Zhang~\emph{et al.}~\cite{ijcai2021p516} and Zhang~\emph{et al.}~\cite{277160} both utilized graph autoencoders to reconstruct the adjacency matrix. The former approach is applicable when the GNN model is fully accessible, while the latter one is employed when a set of public embeddings is available. Duddu~\emph{et al.}~\cite{10.1145/3448891.3448939} used an encoder-decoder framework to reconstruct the target graph. Similar to the previous work, this approach also requires a set of public embeddings and necessitates the incorporation of prior knowledge.  For GRAs with malicious data, Tian~\emph{et al.}~\cite{10386501} performed GRAs by modifying the graph structure. However, this method alters the graph structure without taking into account the concealment of the attack.

\subsection{Poisoning Attacks against GFL Scenarios}
Poisoning attacks refer to deliberately designed small perturbations or modifications that can trick a GNN model into generating incorrect classifications or predictions on input samples. This manipulation of data can provide attackers with additional information to exploit, thereby increasing the risk of compromising a victim's privacy. According to the poison targets, poisoning attacks can be categorized into several types, i.e., modify node features~\cite{bose2019generalizable, takahashi2019indirect, 9745270}, those that alter edges~\cite{bojchevski2019adversarial, xu2019topology, ma2019attacking, zang2020graph}, those that add nodes~\cite{fang2024gani}, and those that rewire connections~\cite{chen2020mga}. For instance, Avishek~\emph{et al.}~\cite{bose2019generalizable} proposed a unified encoder-decoder framework (DAGAER) to generate a complete distribution of poisoned samples, effectively creating a variety of attacks for a single given input. Zang~\emph{et al.}~\cite{zang2020graph} defined anchor nodes that can corrupt a trained graph neural network by flipping the edges of any targeted victim. Fang~\emph{et al.}~\cite{fang2024gani} introduced a global attack strategy via node injection (GANI), which is designed with a comprehensive consideration of an unobtrusive perturbation setting across both structural and feature domains.

The objective of these methods is to alter the prediction outcomes of the primary task through data manipulation. More importantly, the potential risks of privacy leakage associated with these methods have not been investigated.

\section{Preliminaries}
In this section, we briefly introduce the definitions of graph data, GNN models and GFL. We also define the threat model of DMan4Rec. For convenience, the definitions of important symbols used are listed in are given in Appendix~A.

\subsection{Graph and GNN Models}
A graph with $N$ nodes can be represented as $G = \{ V,E,X\} $, where $V = \{ {v_1},...,{v_N}\}$ is the  set of nodes $|V| = N$, ${e_{i,j}} =  < {v_i}{\rm{,}}{v_j} >  \in E$ indicates that there is a link between node ${v_i}$ and ${v_j}$, and ${X} \in {^{N \times L}}$ represents a feature matrix with feature dimension $L$.

The graph dataset $D = \{ G,X,Y\}$ which consists of subgraphs, node features, and labels, is utilized for GNN training and validation. After the training process, the GNN model $f$ is produced, where the model output $f(u)$ represents the posterior probability of node $v_i$ of the class. The primary GNN architectures for node classification include graph convolutional networks (GCN)~\cite{kipf2016semi}, graph sampling and aggregation (GraphSAGE)~\cite{10.5555/3294771.3294869}, and graph attention networks (GAT)~\cite{velivckovic2017graph}. These models employ distinct neural network architectures and learn to aggregate feature information from the node's neighborhood. Their receptive fields are constrained by the model's depth.

\subsection{GFL Scenarios}
In GFL, there are $k$ local clients $C = \{ {C_1}, \ldots ,{C_k}\} $ and a server $S$. The local models have independent model parameters $\Theta  = \{ \theta _c^1, \ldots ,\theta _c^k{\rm{\} }}$, and the server has a top model with parameters ${\theta _t}$. Each client trains an independent graph neural network model using its local data and jointly trains the server model. The subgraph saved in each client ${C_i}$ is horizontally split from the entire underlying graph. This partitioning results in a loss of connectivity due to data isolation. Strictly speaking, there is some overlap, represented as $A \Rightarrow \{ {A^{(k)}}\}$. Each client possesses a subgraph ${G_i}$ within its local dataset ${D_i} = \{ {G_i},X,Y\} $ with the same features. Consequently, the clients share the same feature and label space, although the node ID spaces differ. The global GNN model performs the node classification task:
\begin{equation}
    \mathop {\min }\limits_{{\theta _t}} \frac{{{N_k}}}{N}\sum\limits_{1}^k {{f_k}} ({\theta _t})
\end{equation}

\begin{equation}
    {f_k}({\theta _t}) = {\rm{{\cal L}}}(H({X^{(k)}},{A^{(k)}},{\theta _t}),{Y^{(k)}})
\end{equation}

GFL scenarios are prevalent in the real world. For instance, in online social applications, each user maintains a local social network, and the collective social networks of numerous users form the potential entirety of the human social network. Developers can create friend recommendation algorithms based on these horizontal internal graphs to protect users' social privacy.

\subsection{Threat Model}
\textbf{Scenario.} In GFL, there are $k$ local clients $C=\{ {C_1}, \ldots ,{C_k}\} $ and a server $S$. Each client trains a local model using its own data and uploads the gradient information to the server. During the training process of GFL, the server performs backpropagation, sending updated gradient information back to each client. In this context, it is assumed that one of the clients is a malicious client, and others are benign.

\vspace{0.07cm}
\textbf{Attacker's knowledge.} In GFL, clients share both feature space and label space. Since the attacker is assumed to be a malicious client, their access to the global GNN is limited, as they can only obtain the posterior probabilities of the nodes of interest through queries. The attacker possesses their own local subgraphs, which can be manipulated by either altering node features or modifying subgraph structures to enhance graph inference.

\vspace{0.07cm}
\textbf{Attack goal.} The attacker's objective is to query the trained GNN model to reconstruct the original graph, thereby gaining access to the private data of other clients. This private data includes link information, attribute information, and other sensitive details. Carrying out this process does not interfere with the client's training of the global model, and therefore does not have an impact on the results of the main task's performance, ensuring that the attack is stealthy.

\vspace{0.07cm}
\textbf{Defender goal.} For the benign client as the denfender, the objective is to prevent a malicious attacker from querying the trained GNN model to infer the target graph, all while maintaining the performance of the primary task.

\section{Methodology}
To investigate the privacy leakage vulnerabilities in GFL, we propose a data manipulation aided reconstruction attack (DMan4Rec) from the perspective of a malicious client. The main goal is to execute a graph reconstruction attack by manipulating the data of a malicious client during GFL process, while avoiding the degradation of the main task's performance and ensuring the attack remains concealed. In order to achieve this goal, DManRec is implemented by two stages, i.e., training and inference. In the training stage, the attacker modifies the objective function by incorporating a loss term based on label smoothing, a loss term based on connected nodes, and a loss term based on non-connected nodes. This adjustment updates the node features to create a manipulated graph for training malicious models. Simultaneously, benign clients also train local models. Subsequently, the clients upload their respective model parameters to the server, which aggregates them to update the global model for subsequent main tasks. And in the inference stage, the attacker reconstructs the target graph by querying the node posterior probabilities, which serve as input for the attack model.

\begin{figure*}[htbp!]
	\centering
	\includegraphics[width=0.8\linewidth]{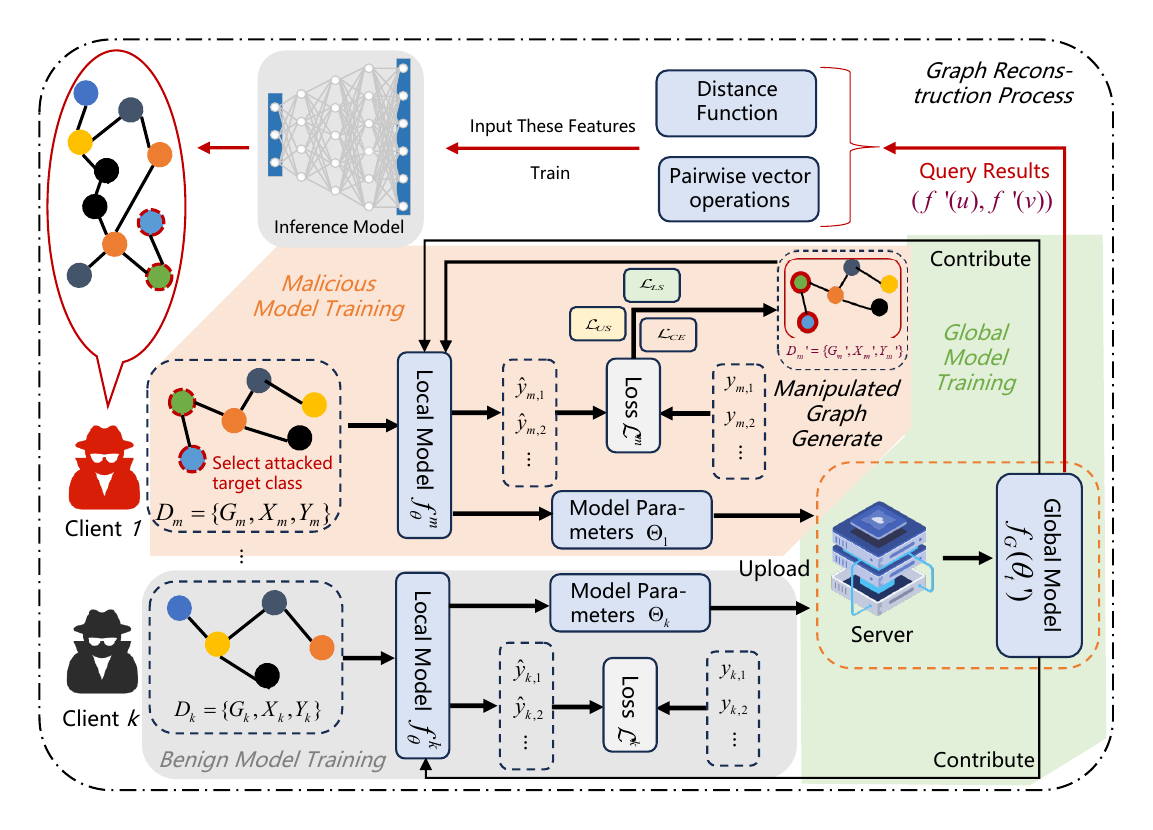}\\
	\caption{An illustration of the graph reconstruction attack on GFL based on data manipulation. }
	\label{fig2}
\end{figure*}

The framework of DMan4Rec is shown in Fig.~\ref{fig2}. In this section, we will detail DMan4Rec in terms of manipulated graph training, malicious model training, and the graph reconstruction process.

\subsection{Manipulated Graph Training}
Assume that the malicious client has the original local dataset ${D_m} = \{ {G_m},{X_m},{Y_m}\}$, which contains the subgraph ${G_m}$ owned by the client, and trains the local model $f_\theta ^m$. To enhance the adversarial robustness of the malicious model, we opt for the cross-entropy term as the regularization term in the loss function:
\begin{equation}
    {{\rm{{\mathcal L}}}_{CE}} =  - \sum\nolimits_{n = 1}^n {\log ({{\hat y}_k})} {y_k},
\end{equation}
where $n$ is the number of node categories, ${\hat y_k}$ is the predicted label, and ${y_k}$ is the true label.

\vspace{0.1cm} 
Then, to prevent the model from overfitting and to enhance its generalization ability, we introduce the label smoothing method, which modifies the label ${y_k}$ to obtain:
\begin{equation}
    {y_k}' = {y_k}(1 - \varepsilon ) + \varepsilon /{{\rm{{\mathcal L}}}_{CE}},
\end{equation}
where $\varepsilon$ is the label-smoothing hyperparameter.

\vspace{0.1cm}
Furthermore, to generate manipulated graphs that encourage the trained GNN model to focus more on adjacency, it is essential to enhance the similarity of outputs for connected nodes while promoting dissimilarity for non-connected nodes. To accomplish this, two regularization terms are introduced:
\begin{equation}
    {{\rm{{\mathcal L}}}_{LS}} =  - {\sum\limits_{(u,v) \in E} {(f_\theta ^m(u) - f_\theta ^m(v))} ^2}
\end{equation}

\begin{equation}
    \mathcal{L}_{U S}=-\sum_{\substack{v \in \in v, u+\theta \\(u, v) \notin E}}\left(1-\cos \left(f_{\theta}^{m}(u), f_{\theta}^{m}(v)\right)\right)^{2}
\end{equation}

By calculating the Euclidean distance ${{\rm{{\mathcal L}}}_{LS}}$ of the posterior probabilities on two connected nodes, we can identify node features that minimize the distance between these connected nodes. As cosine similarity is constrained, to prevent the dissimilarity term from becoming excessively large, we opt to identify node features that decrease the similarity between unconnected nodes by computing the cosine similarity ${{\rm{{\mathcal L}}}_{US}}$ between them.

\vspace{0.1cm} 
In all, the objective function to add on node features is expressed as follows:
\begin{equation}
    {\rm{{\mathcal L}}} = \alpha {{\rm{{\mathcal L}}}_{LS}} + \beta {{\rm{{\mathcal L}}}_{US}} + \lambda {{\rm{{\mathcal L}}}_{CE}},
\end{equation}
where $\alpha $,  $\beta$, $\lambda $ are the regularization positive coefficients.

\vspace{0.1cm} 
Then the projected gradient descent algorithm can be used to update the node features:
\begin{equation}
    {x_{n + 1}} = {x_n} + \eta \nabla {\rm{{\mathcal L}}},
    \label{q8}
\end{equation}
where $\eta $ is the size of iteration steps.

\vspace{0.1cm} 
Therefore, the manipulated graph dataset can be represented as ${D_m}' = \{ {G_m}',{X_m}',{Y_m}'\} $.

\subsection{Malicious Model Training}
The global GNN model uses all the parameters uploaded by the client, including the toxic parameters uploaded by the malicious client, to execute the node classification task:
\begin{equation}
    {f_G}({\theta _t}') = {\rm{{\mathcal L}}}(H({X^{(k)}}',{A^{(k)}}',{\theta _t}'),{Y^{(k)}}'),
    \label{q9}
\end{equation}
where ${X^{(k)}}',{A^{(k)}}' \in D \cup {D_m}'$.

\subsection{Graph Reconstruction Process}
The malicious client queries the global model to obtain the posterior probability $({f_{}}'(u),{f_{}}'(v))$ of the node pair, that is, the node posterior matrix $H$ is obtained, and a supervised model is trained using it. To mitigate the sensitivity of the distance function, we initially compute the 8 distance functions~\cite{2005.02131} of $({f_{}}'(u),{f_{}}'(v))$.

\textbf{Cosine Distance.} The cosine distance is defined by measuring the difference in direction between the posterior probability of output nodes $u$ and $v$:
\begin{equation}
    Cosine{\rm{ }}Distance = 1 - \frac{{{f_{}}'(u) \cdot {f_{}}'(v)}}{{{{\left\| {{f_{}}'(u)} \right\|}_2}{{\left\| {{f_{}}'(v)} \right\|}_2}}}
\end{equation}
The closer the cosine is to 1, the more aligned they are, the more similar they are.

\textbf{Euclidean Distance.} The absolute distance between the output nodes $u$ and $v$ is defined as follows: 
\begin{equation}
    Euclidean{\rm{ }}Distance = {\left\| {{f_{}}'(u) - {f_{}}'(v)} \right\|_2}
\end{equation}
The smaller the value, the more similar they are. 

\textbf{Correlation Distance.} Correlation distance is defined as:
\begin{equation}
\begin{split}
    Correlation{\rm{ }}Distance = \\ 1 - \frac{{({f_{}}'(u) - {{\bar f}_{}}'(u)) \cdot ({f_{}}'(v) - {{\bar f}_{}}'(v))}}{{{{\left\| {({f_{}}'(u) - {{\bar f}_{}}'(u))} \right\|}_2}{{\left\| {({f_{}}'(v) - {{\bar f}_{}}'(v))} \right\|}_2}}},
\end{split}
\end{equation}
where the subtractor represents the correlation coefficient, which measures the correlation degree of the output nodes $u$ and $v$. The value range of the correlation coefficient is $[-1,1]$. The greater the absolute value of the correlation coefficient, the smaller the distance function value, and the higher the correlation between the output of node pairs. When the two are linearly correlated, the correlation coefficient takes the value of 1 (positive linear correlation) or -1 (negative linear correlation). 

\textbf{Chebyshev Distance.} Chebyshev distance emphasizes the maximum difference between two output nodes in any single dimension, rather than the sum of all dimensional differences. This metric offers a method for measuring multidimensional differences. A smaller distance indicates a greater similarity between the output node pair.
\begin{equation}
    Chebyshev{\rm{ }}Distance = {\max _i}\left| {{f_i}'(u) - {f_i}'(v)} \right|
\end{equation}

\textbf{Braycurtis Distance.} The braycurtis distance is not strictly a distance measure, because it does not satisfy the triangle inequality. but it is a very practical measure of similarity. For the output nodes $u$ and $v$, it can be calculated by the following formula:
\begin{equation}
    Braycurtis{\rm{ }}Distance = \frac{{\sum {_i\left| {{f_i}'(u) - {f_i}'(v)} \right|} }}{{\sum {_i\left| {{f_i}'(u) + {f_i}'(v)} \right|} }}
\end{equation}
The distance ranges from 0 to 1, where 0 means the two nodes are exactly the same and 1 means they are completely different.

\textbf{Manhattan Distance.} The Manhattan distance, also known as the city block distance, is expressed as follows:
\begin{equation}
    Manhattan {\rm{ }}Distance = \sum {_i\left| {{f_i}'(u) - {f_i}'(v)} \right|} ,
\end{equation}
A smaller distance indicates a greater similarity between the outputs node pair. This is suitable for high-dimensional data and is not susceptible to outliers in individual dimensions.

\textbf{Camberra Distance.} The camberra distance is a numerical measure of the difference between two vectors, especially for non-negative numerical data, where for any two output nodes the distance function is defined as follows:
\begin{equation}
    Camberra{\rm{ }}Distance = \sum {_i\frac{{\left| {{f_i}'(u) - {f_i}'(v)} \right|}}{{\left| {{f_i}'(u)} \right| + \left| {{f_i}'(v)} \right|}}} 
\end{equation}
The smaller the distance value, the higher the similarity.

\textbf{Sqeuclidean Distance.} The sqeuclidean distance represents the square Euclidean distance between two outputs of node pairs ${f'(u), f'(v)}$ as follows:
\begin{equation}
    Sqeuclidean{\rm{ }}Distance = \left\| {f'(u) - f'(v)} \right\|_2^2
\end{equation}
The smaller the distance value, the greater the similarity between two output nodes.

Moreover, four entropy features are calculated: $ \left( {{f_i}'(u) + {f_i}'(v)} \right)/2$ (Average),$ \left| {{f_i}'(u) - {f_i}'(v)} \right|$ (Weighted-L1), $ {f_i}'(u) \cdot {f_i}'(v)$ (Hadamard), ${\left| {{f_i}'(u) - {f_i}'(v)} \right|^2}$ (Weighted-L2). These indicators are used to obtain feature vectors for training the attack model. The attack model uses a multi-layer perceptron (MLP) to predict ${A_{uv}}$, obtaining the entire adjacency matrix $A$.








However, multi-layer perceptrons (MLPs) often struggle to adequately capture the complex dependencies between similar features. Furthermore, the manipulated graph introduces additional complexities, such as the similarity of connected nodes and the dissimilarity of non-connected nodes. To address these challenges, we propose a multi-head self-attention mechanism based on MLPs. Specifically, we firstly create a novel attack model based on self-attention mechanism by obtaining weights and biases from the first layer of MLP, mapping the similarity features of the input to the embedded feature dimension, transforming each input vector through a linear layer and multiple attention calculations, and ultimately outputting the predicted results through a fully connected layer to generate the complete adjacency matrix.
This mechanism effectively utilizes information by selectively focusing on various aspects of similar input features.


The complete algorithm of DMan4Rec is as shown in Appendix B. Besides, theoretical analysis and Time Complexity Analysis are also given in Appendix C and D.

\section{Experiments}
In this section, we comprehensively evaluate the proposed DMan4Rec. In particular, we aim to answer the following research questions (RQs):
 \begin{itemize}
\item \textbf{RQ1:} Can DMan4Rec achieve the SOTA attack performance in white-box scenario?
\item \textbf{RQ2:} Can DMan4Rec outperform the SOTA baselines in a black-box scenario by transferable GNN models?
\item \textbf{RQ3:} How DMan4Rec maintains stealthy?
\item \textbf{RQ4:} How effective is DMan4Rec in ablation study?
\item \textbf{RQ5:} Can DMan4Rec still beat defensive GFLs?
\item \textbf{RQ6:} How sensitive DMan4Rec is?
\end{itemize}

\subsection{Datasets}
In order to verify the scalability of DMan4Rec across various datasets, we selected four publicly available datasets of differing sizes for evaluation: Cora~\cite{McCallum2000AutomatingTC} , Citeseer~\cite{McCallum2000AutomatingTC}, Amazon Photo~\cite{DBLP:journals/corr/abs-1811-05868}, and Amazon Computer~\cite{DBLP:journals/corr/abs-1811-05868}. Detailed statistics for these datasets are presented in Appendix H. 


\subsection{Evaluation Metrics}
In order to verify the effectiveness of DMan4Rec, we use the area under the ROC curve (AUC) and inference precision (Precision) as measures of the effectiveness of GRAs which is consistent with previous work~\cite{Kipf2016VariationalGA}. Classification accuracy (Acc)~\cite{Kipf2016VariationalGA} is employed to assess the performance of node classification tasks. In addition, we utilize the AUC of the similarity between connected and unconnected nodes (AUC-CUS) to evaluate the stealthiness of DMan4Rec. Details are provided in the Appendix E.

\subsection{GNN models}
In order to prove that DMan4Rec is effective in the structure of GFL based on different GNN models, three GNN models including GCN~\cite{DBLP:conf/iclr/KipfW17}, GraphSAGE~\cite{10.5555/3294771.3294869} and GAT~\cite{DBLP:conf/iclr/VelickovicCCRLB18} are used as the local models of the participants. Details are provided in the Appendix F.

\subsection{Baselines}
Since this is the first study of data manipulation aided reconstruction attack on GFL, four GRA methods and one GRA method with data manipulation under the centralized GNN setting are selected as comparison algorithms. In order to make the attack methods transferable to GFL, the comparison algorithms uniformly follow the GFL setting. In order to better distinguish the prior knowledge comparison of these methods, we classify them as shown in TABLE~\ref{t2}, where a hollow circle indicates that the knowledge can be directly acquired, a solid circle indicates that the knowledge is unknown, and a semi-solid circle indicates that the knowledge cannot be directly acquired but can be constructed by oneself, which is similar to the target knowledge. The comparison algorithms are briefly described as follows:

\begin{table}[htbp]
\setlength{\abovecaptionskip}{0.05cm}\setlength{\belowcaptionskip}{-0.2cm}\setlength{\abovecaptionskip}{0.1cm}
	\centering
	\caption{DIFFERENT METHODS WITH DIFFERENT ACCESSIBLE KNOWLEDGE}
	\scalebox{0.85}{
		\begin{tabular}{c|ccccc}
			\toprule \hline
			Methods   & $X$  & $Y$  & ${A^*}$ & Global model &$E$
\\ \hline
			GraphMI & \emptycirc &\emptycirc  &\fullcirc  &\emptycirc &\fullcirc \\
			QPLGE &\fullcirc  &\fullcirc   &\halfcirc &\emptycirc &\emptycirc \\
			IAGNN &\fullcirc  &\fullcirc  &\halfcirc &\emptycirc &\emptycirc \\
			GRecon &\emptycirc &\emptycirc &\emptycirc  &\fullcirc  &\fullcirc  \\
   GCP-GRecon &\emptycirc &\emptycirc &\emptycirc  &\fullcirc  &\fullcirc  \\
   DMan4Rec &\emptycirc &\emptycirc &\emptycirc  &\fullcirc  &\fullcirc  \\
   \hline \bottomrule
		\end{tabular}} 
  \label{t2} 
\end{table}

\begin{table*}[htbp!]
    \centering
    
    \caption{COMPARISON OF AUC AND AP OF DIFFERENT ATTACKS AND PERFORMANCE OF NODE CLASSIFICATION TASK UNDER GFL SETTING}
    \scalebox{0.68}{
    \begin{tabular}{c|c|c|ccc|ccc|ccc}
    \toprule\hline
         \multirow{2}{*}{Dataset} & \multirow{2}{*}{Method Classification} & \multirow{2}{*}{Methods} & \multicolumn{3}{c|}{GraphSAGE} & \multicolumn{3}{c|}{GCN} & \multicolumn{3}{c}{GAT} \\ 
        \cline{4-12} 
         &  &   & AUC (\%) & Precision (\%) & Acc (\%) & AUC (\%) & Precision (\%) & Acc (\%) & AUC (\%) & Precision (\%) & Acc (\%) \\ \hline
        \multirow{6}{*}{Cora} & \multirow{4}{*}{GRA only} & GraphMI & 50.00 & 50.00 & 76.11 & 50.00 & 50.00 & 79.30 & 50.00 & 50.00 & 78.59 \\ 
        ~ & ~ & QPLGE & 55.07 & 50.83 & 76.70 & 59.58 & 64.49 & 79.56 & 56.16 & 53.64 & 77.70 \\ 
        ~ & ~ & IAGNN & 85.11 & 83.20 & 76.70 & 86.45 & 79.01 & 79.30 & 84.34 & 83.37 & 77.70 \\ 
        ~ & ~ & GRecon & 88.51 & 90.74 & 86.72 & 87.29 & 79.41 & 89.15 & 86.29 & 86.86 & 88.02 \\ 
        \cline{2-12}
         & \multirow{2}{*}{DM-GRA} & GCP-GRecon & 95.78 & 95.43 & 84.67 & 87.31 & 82.90 & 87.44 & 92.08 & 91.99 & 86.45 \\ 
        ~ & ~ & DMan4Rec & \textbf{96.65} & \textbf{96.78} & \textbf{88.75} & \textbf{88.61} & \textbf{86.39} & \textbf{88.26} & \textbf{93.54} & \textbf{92.84} & \textbf{87.26} \\ \hline
        \multirow{6}{*}{Citeseer} & \multirow{4}{*}{GRA only} & GraphMI & 50.00 & 50.00 & 63.48 & 50.00 & 50.00 & 64.30 & 50.00 & 50.00 & 72.13 \\ 
        ~ & ~ & QPLGE & 54.39 & 62.52 & 62.80 & 51.33 & 58.18 & 64.25 & 54.64 & 50.94 & 62.10 \\ 
        ~ & ~ & IAGNN & 84.99 & 79.25 & 62.80 & 84.80 & 80.11 & 64.30 & 88.70 & 86.96 & 62.10 \\ 
        ~ & ~ & GRecon & 92.90 & 87.50 & 88.71 & 96.98 & 95.76 & 89.58 & 97.77 & 95.72 & 99.97 \\ 
        \cline{2-12} & \multirow{2}{*}{DM-GRA} & GCP-GRecon & 96.87 & 95.43 & 86.67 & 98.78 & 98.67 & 85.98 & 98.72 & 97.89 & 96.98 \\ 
        ~ & ~ & DMan4Rec & \textbf{97.76} & \textbf{96.71} & \textbf{87.55} & \textbf{99.37} & \textbf{99.41} & \textbf{87.07} & \textbf{99.72} & \textbf{99.75} & \textbf{97.18} \\ \hline
        \multirow{6}{*}{Amazon Photo} & \multirow{4}{*}{GRA only} & GraphMI & 50.00 & 50.00 & 70.25 & 50.00 & 50.00 & 73.56 & 50.00 & 50.00 & 72.15 \\ 
        ~ & ~ & QPLGE & 57.63 & 55.78 & 70.36 & 50.87 & 58.01 & 73.78 & 51.21 & 56.78 & 72.52 \\ 
        ~ & ~ & IAGNN & 87.46 & 82.52 & 70.45 & 84.12 & 82.01 & 74.35 & 86.46 & 80.23 & 72.56 \\ 
        ~ & ~ & GRecon & 95.35 & 94.58 & 93.35 & 85.28 & 82.12 & 93.73 & 87.40 & 80.87 & 93.56 \\ 
        \cline{2-12} & \multirow{2}{*}{DM-GRA} & GCP-GRecon & 97.89 & 96.01 & 91.65 & 89.45 & 83.78 & 91.77 & 89.01 & 84.21 & 92.87 \\ 
        ~ & ~ & DMan4Rec & \textbf{98.22} & \textbf{96.98} & \textbf{92.23} & \textbf{89.64} & \textbf{83.80} & \textbf{92.19} & \textbf{89.07} & \textbf{84.41} & \textbf{93.01} \\ \hline
        \multirow{6}{*}{Amazon Computer} & \multirow{4}{*}{GRA only} & GraphMI & 50.00 & 50.00 & 74.56 & 50.00 & 50.00 & 77.26 & 50.00 & 50.00 & 79.15 \\ 
        ~ & ~ & QPLGE & 53.67 & 50.23 & 74.78 & 50.65 & 50.13 & 77.46 & 59.37 & 60.12 & 79.32 \\ 
        ~ & ~ & IAGNN & 83.23 & 80.67 & 74.35 & 73.98 & 70.56 & 77.45 & 82.13 & 79.45 & 79.10 \\ 
        ~ & ~ & GRecon & 91.73 & 87.88 & 85.83 & 85.72 & 78.25 & 89.48 & 89.94 & 85.16 & 92.81 \\ 
        \cline{2-12} & \multirow{2}{*}{DM-GRA} & GCP-GRecon & 92.12 & 89.56 & 82.84 & 87.01 & 81.12 & 85.87 & 90.87 & 86.99 & 91.89 \\ 
        ~ & ~ & DMan4Rec & \textbf{92.20} & \textbf{89.77} & \textbf{89.32} & \textbf{88.18} & \textbf{82.09} & \textbf{87.04} & \textbf{91.48} & \textbf{87.68} & \textbf{92.26} \\ \hline\bottomrule
    \end{tabular}}
    \label{tab:S1}
\end{table*}

\begin{itemize}
    \item  GraphMI~\cite{ijcai2021p516}. Infer the true graph structure by initializing the adjacency matrix and continuously accessing the target model, while minimizing the difference between the obtained graph embedding and the true graph embedding. Since this attack method is performed under the white-box setting, in order to unify the GFL setting, the malicious client can only query the posterior probability nodes and cannot obtain the global model parameters.
    \item QPLGE~\cite{10.1145/3448891.3448939}. The attacker can access the node embeddings of the subgraph, train the encoder-decoder model, and reconstruct the target graph from the publicly released embeddings.
    \item IAGNN~\cite{277160}. The attacker trains a graph autoencoder model. After the graph autoencoder is trained, its decoder is used as the attack model to reconstruct the target graph from the publicly released embeddings.
    \item GRecon~\cite{10288405}. The attention model trained using entropy features and distance function features is used as the attack model to reconstruct the target graph.
    \item GCP-GRecon~\cite{10386501}. By modifying the links between malicious nodes, a manipulable graph is generated as the input of the attack model.
\end{itemize}

\subsection{Parameters Settings}
Our experiments have the following settings: the size of hidden layer of GNN model was 16, unless otherwise specified; to train the target model, the cross-entropy loss and the Adam optimizer with the learning rate set to 0.01 were used; the number of PGD steps was 100; all the models were trained for 100 epochs. To mitigate non-determinism, we repeated the experiment for 3 times and reported the average results.

\vspace{0.08cm}
Our method is implemented in Python based on the Pytorch framework. All the experiments are conducted on a server machine equipped with i7-7700K 3.5GHzx8 (CPU), TITAN Xp 12GiB (GPU), 16GBx4 memory (DDR4), and Ubuntu 18.04 (OS).

\subsection{White-box Attack of DMan4Rec (RQ1)}
When reporting the results, we focus on whether DMan4Rec achieves SOTA attack performance in white-box scenarios.

\textbf{Implementation Details.} 
Under the GFL scenario, we implemented DMan4Rec on four datasets using three GNN models and compared it with the other data manipulation method. We report AUC and  precision scores for attacks and Acc scores for node classification task. The results are shown in TABLE~\ref{tab:S1}, with the best attack performance in bold. 

\textbf{Results and Analysis.} 
From TABLE~\ref{tab:S1}, it is observed that DMan4Rec achieves the best performance, with an AUC score of 99.75\% and a precision score of 97.18\% on the Citeseer dataset using the GAT model. Compared with the data manipulation method (GCP-GRecon) of modifying the link, the performance is improved by about 1\%.


Furthermore, for the main task performance, the Acc score of DMan4Rec remains relatively stable compared to GRA only, with an average error range of about 0.76\%. The presence of this slight deviation also ensures the concealment of the attack, as each Acc value for the server is obtained through retraining. Compared with GCP-GRecon, when the attack performance is not much different, it cannot guarantee the stability of the main task performance, because the means of data manipulation is to modify the link, which is easy to reduce the main task performance.


\subsection{Black-box Attack of DMan4Rec (RQ2)}
In this part, we focus on whether DMan4Rec can successfully launch an attack in black-box scenarios using transferable GNN models. Specifically, we aim to investigate the effectiveness of the attack executed by the attacker utilizing a model that differs from the global model, particularly when the attacker remains unknown to the global model.

\begin{figure*}[htbp!]
	\centering
	\includegraphics[width=1\linewidth]{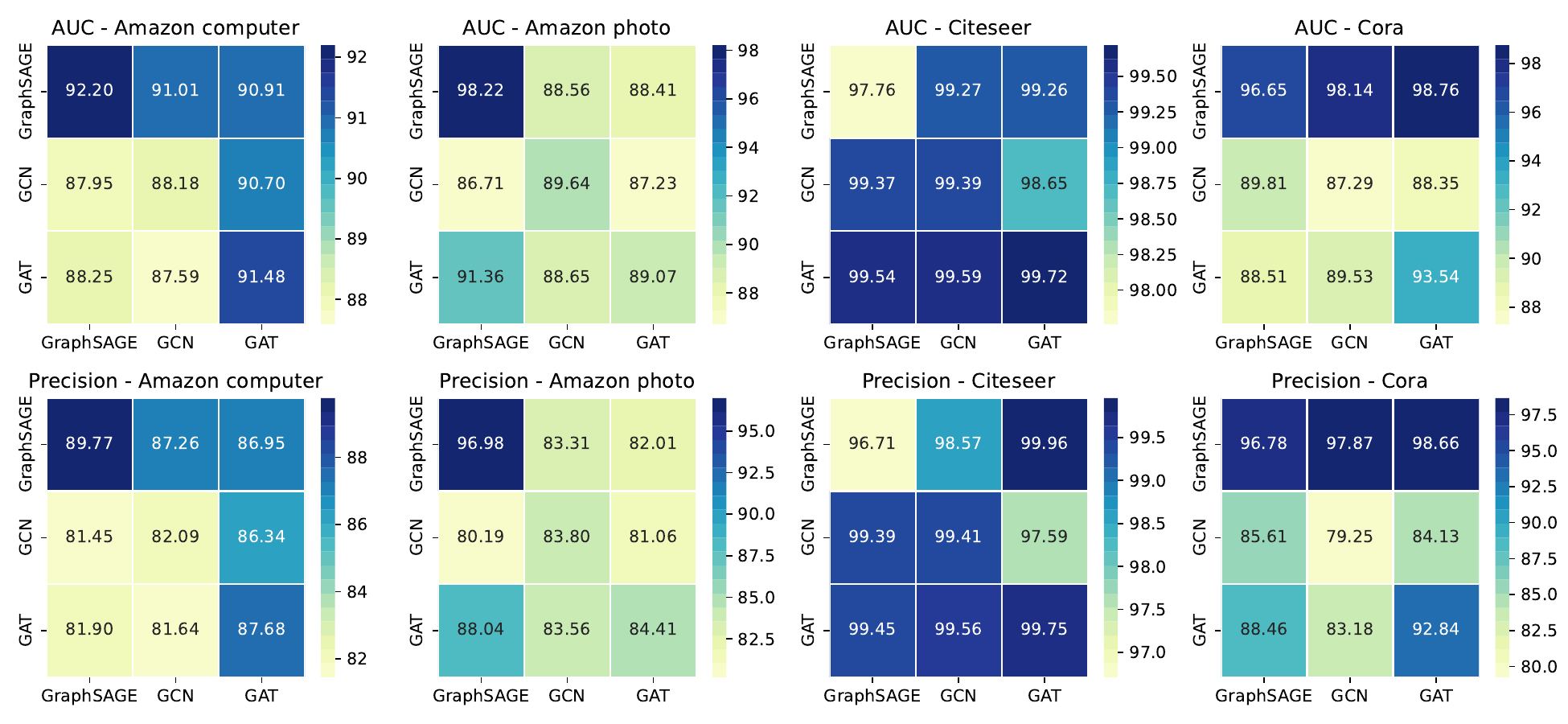}\\
	\caption{Black-box attack performance of DMan4Rec on different datasets.}
	\label{fig5}
\end{figure*}

\textbf{Implementation Details.}
We conducted black-box attack experiments on four distinct datasets using three GNN models. As illustrated in Fig.~\ref{fig5}, the vertical axis represents the server model, while the horizontal axis represents the malicious client attack model.

\textbf{Results and Analysis.}
As depicted in Fig.~\ref{fig5}, we observed that the attack's AUC score and Precision score reached up to 99.72\% and 99.96\% respectively, which shows that even without knowledge of the server model structure, the attacker can utilize DMan4Rec to achieve high-performance GRAs. Moreover, when the server model is GraphSAGE, regardless of the malicious client attack model being one of the three models, the impact is superior compared to when the server model is either of the other two. One potential explanation is that GraphSAGE is more proficient in integrating global information, while the local aggregation techniques of GCN and GAT may be more adept at capturing local information and features. For the Citeseer dataset, the effectiveness of using different models for attacking exceeded 97.76\%, hence the distinction in the aforementioned phenomenon is not significant. Intriguingly, on the Citeseer dataset, the highest AUC score and Precision score are achieved when both the malicious client attack model and the server model are GAT. This implies that when the attacker employs GAT as a local model to generate the manipulated graph, the reason may lie in GAT's enhanced generalization in estimating the true boundary of the server model, rendering GAT's manipulated graph more impactful.

\subsection{Attack Concealment (RQ3)}

When reporting the results, we focus on the following aspects: homogeneous distribution before and after data manipulation, AUC-CUS scores of connected and non-connected nodes.

\textbf{Implementation Details.}
We aim to compare the homogeneous distribution before and after data manipulation on two Amazon datasets using three GNN models. The results are presented in Fig.~\ref{fig3}. Additionally, we report the AUC-CUS scores as shown in Appendix G. We select various similarity functions as inputs for the AUC-CUS scores to ensure the applicability of DMan4Rec to different similarity function calculation methods, thereby proving the absolute accuracy of DMan4Rec.

\vspace{0.08cm}
\textbf{Results and Analysis.} 
The node-centered homogeneous distribution shifts between the clean graph and the manipulated graph, with a threshold upper limit, ensuring that the server cannot easily detect malicious nodes. As shown in Fig.~\ref{fig3}, the distribution graphs of the benign graph and the manipulated graph are nearly completely overlapped, indicating that DMan4Rec can effectively preserve homogeneity while executing effective data manipulation. One potential explanation is that we opted to manipulate the node features without altering the entire graph structure, rendering the attack challenging to detect.


\begin{figure*}[htbp]
  \centering
  
    \includegraphics[width=0.31\linewidth]{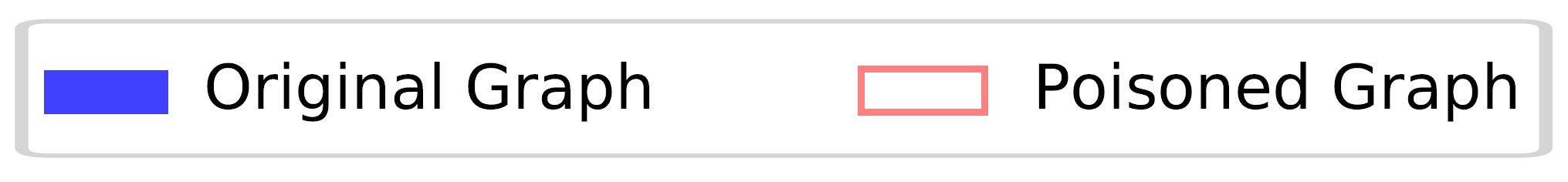}
    
  \subfigure[Amazon Photo dataset using GraphSAGE model]{
    \includegraphics[width=0.31\linewidth]{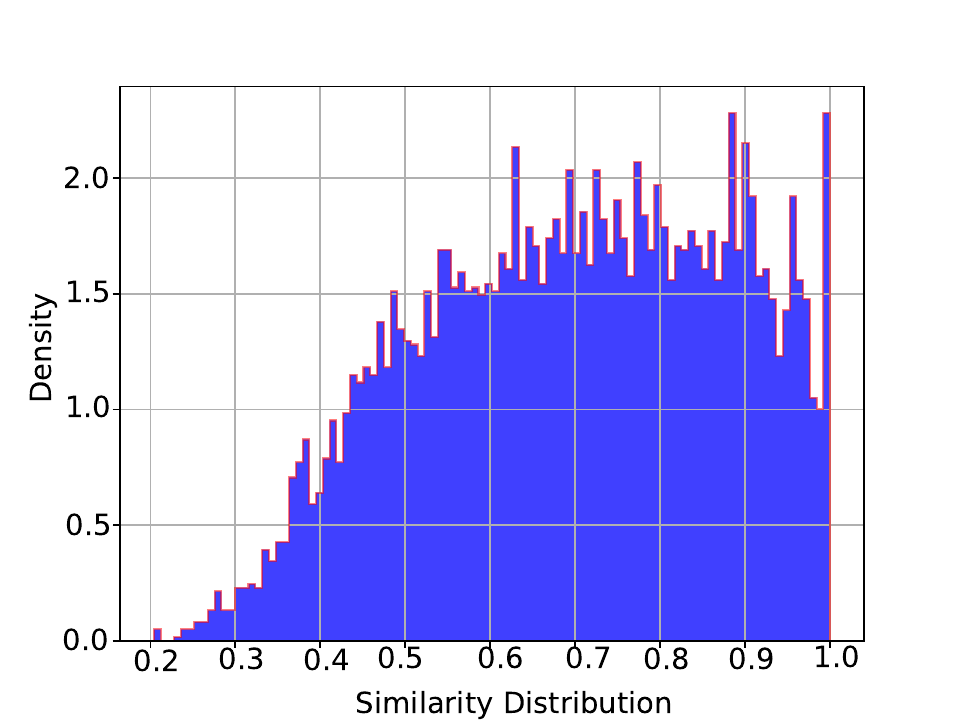}
    } 
  \subfigure[Amazon Photo dataset using GCN model]{
    \includegraphics[width=0.31\linewidth]{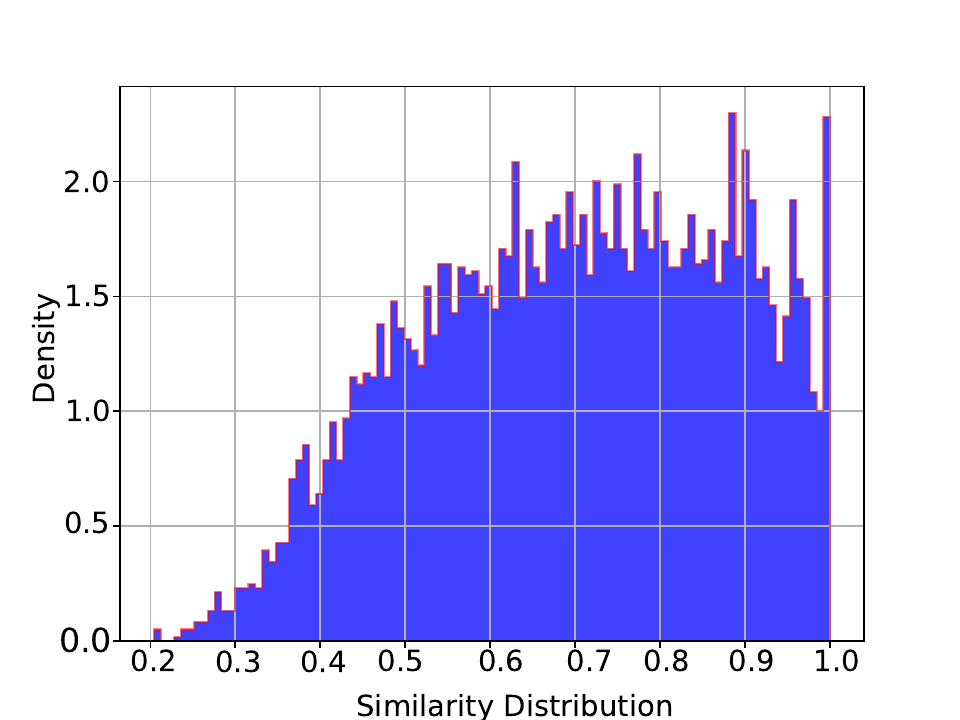}
    } 
  \subfigure[Amazon Photo dataset using GAT model]{
    \includegraphics[width=0.31\linewidth]{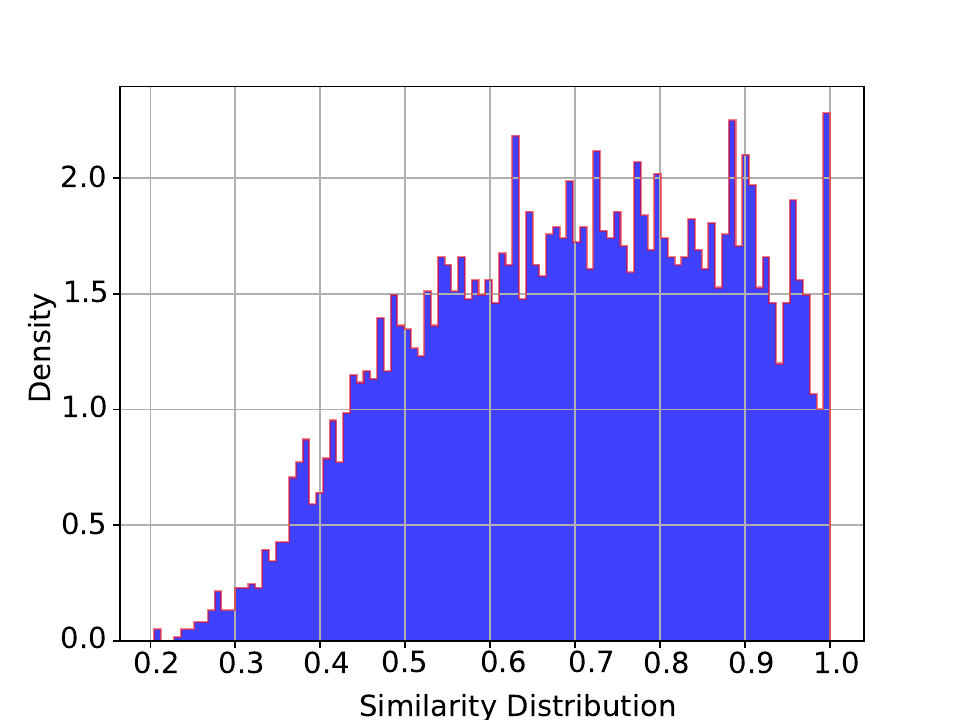}
    }
    \\
    \subfigure[Amazon Computer dataset using GraphSAGE model]{
    \includegraphics[width=0.31\linewidth]{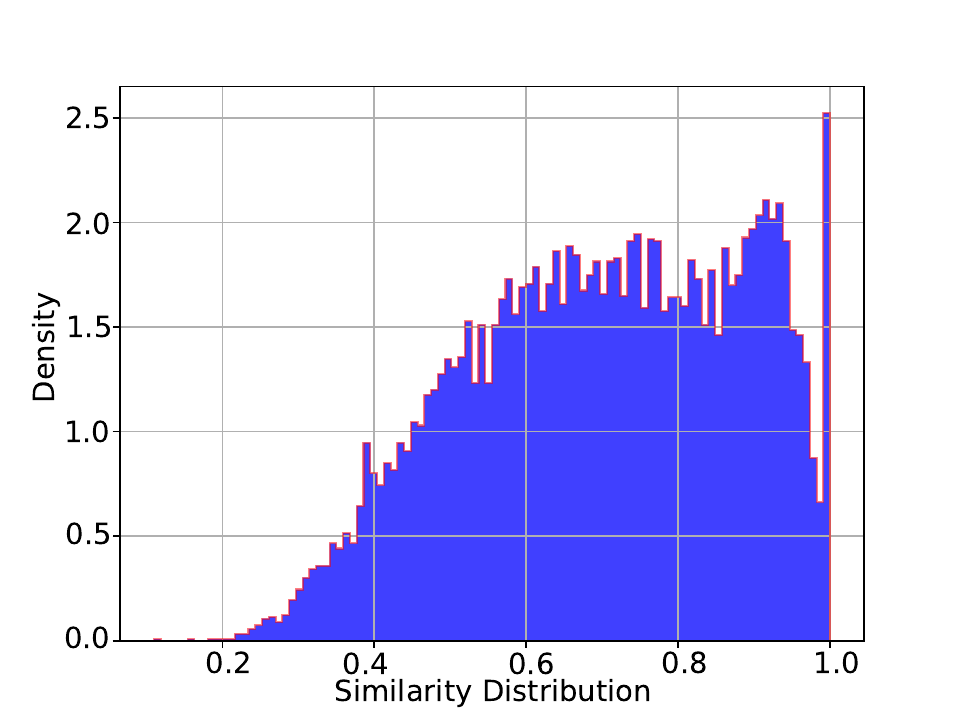}
    } 
  \subfigure[Amazon Computer dataset using GCN model]{
    \includegraphics[width=0.31\linewidth]{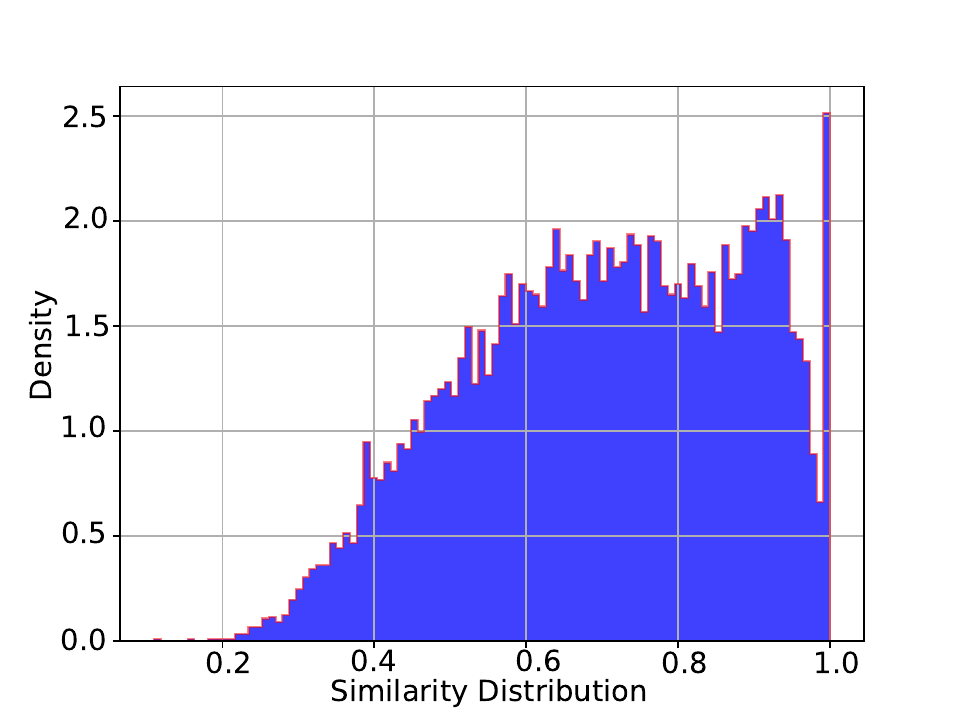}
  }
  \subfigure[Amazon Computer dataset using GAT model]{
    \includegraphics[width=0.31\linewidth]{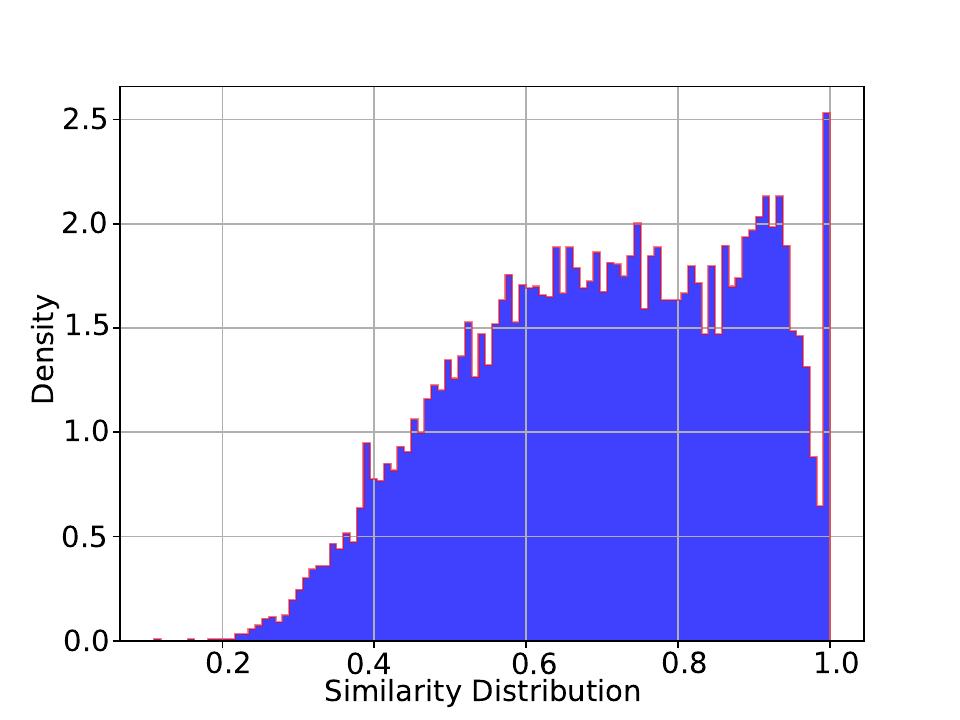}
  }
   \caption{Similarity distribution of two Amazon datasets before and after DMan4Rec poisoning.}
    \label{fig3}
\end{figure*}

\vspace{0.08cm}




\subsection{Ablation Study (RQ4)}
To further explore the effectiveness of data manipulation, we conducted ablation experiments on the manipulation graph generation module.

\textbf{Implementation Details.} Under GFL secnario, we implemented six GRAs including GRA with or without manipulation graph generation module, namely GRA only and DM-GRA. We report AUC and  precision scores for attacks and Acc scores for node classification task on four datasets and three GNN models. The results are shown in TABLE~\ref{tab:S1}, with the best attack performance in bold. 

\textbf{Results and Analysis.} 
From TABLE~\ref{tab:S1}, we can see that DMan4Rec achieves the best performance on all datasets and all models, which proves that DMan4Rec can significantly improve the effectiveness of GRAs.

Among the five methods of GRA, GraphMI has the worst attack effect because the malicious client cannot obtain prior knowledge of the global model parameters. To make up for the lack of prior knowledge, we set the global model parameters to random values so that GraphMI can only reach the level of random guessing. QPLGE and IAGNN methods both require a set of embeddings as prior knowledge. We utilize the local embeddings of the malicious client as input to reconstruct the target graph, enhancing its attack effectiveness compared to GraphMI. The variance lies in the chosen training models: QPLGE opts for an encoder-decoder, whereas IAGNN selects a graph autoencoder (GAE). GAE excels in simulating intricate network relationships among nodes and, through mastering an efficient encoding method for graph structures, it can more effectively retain the graph's structure and attribute information. GRA has the most significant impact among GRAs because this approach does not necessitate prior knowledge. It only requires querying the global model to acquire the input for the attack model to reconstruct the target graph.

Since GRA performs best among the methods of GRA only, we perform data manipulation before GRA. That is, we perform GRA after generating a manipulated graph. 
Compared with the method of GRA only, the attack AUC and Precision is improved by up to 9.2\% and 10.5\% respectively. The additional information (malicious nodes) can be utilized for GRA, increasing the possibility of an attack. Furthermore, for the main task performance, the Acc score of DMan4Rec remains relatively stable compared to GRA only, with an average error range of about 0.76\%. The presence of this slight deviation also ensures the concealment of the attack, as each Acc value for the server is obtained through retraining.


\subsection{Possible Defense against DMan4Rec (RQ5)}
In this section, we focus on whether DMan4Rec can still function under potential defenses. Specifically, we enhance privacy by introducing noise to the model gradients.

\begin{figure*}[htbp]
  \centering
  {
    \includegraphics[width=0.16\linewidth]{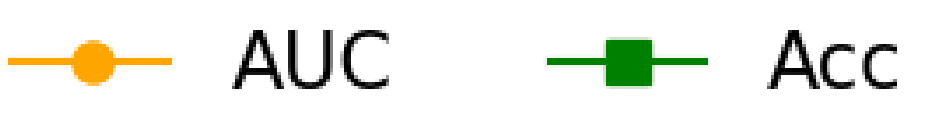}
    } 
    \subfigure[The attack performance of DMan4Rec on Amazon Photo dataset by adding Laplacian noise.]
  {
    \includegraphics[width=0.9\linewidth]{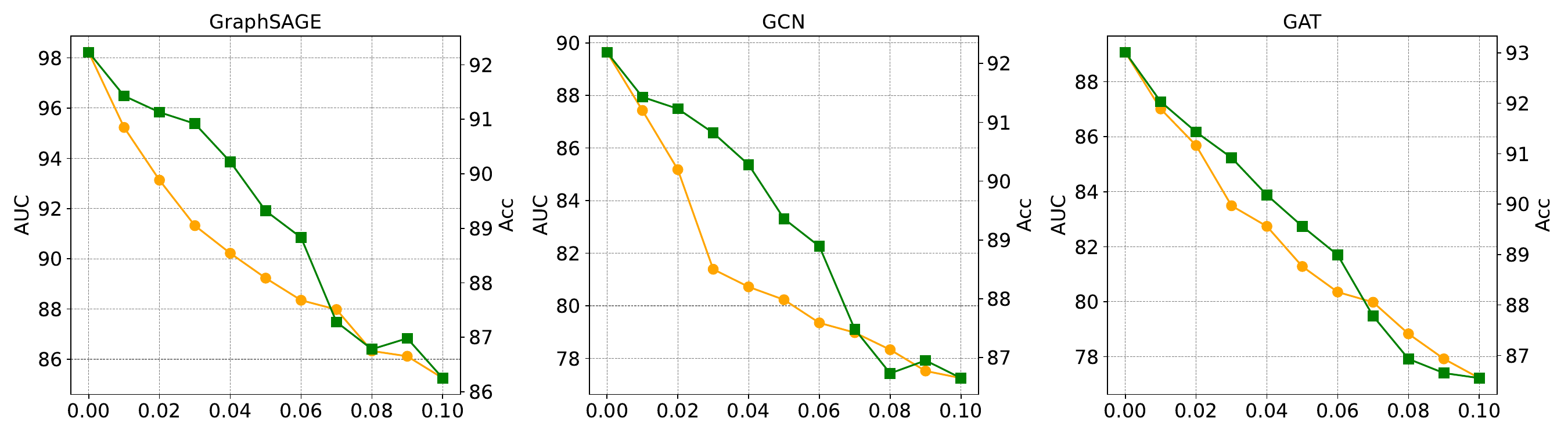}
    } 

    \subfigure[The attack performance of DMan4Rec on Amazon Photo dataset by adding Gaussian noise.]{
    \includegraphics[width=0.9\linewidth]{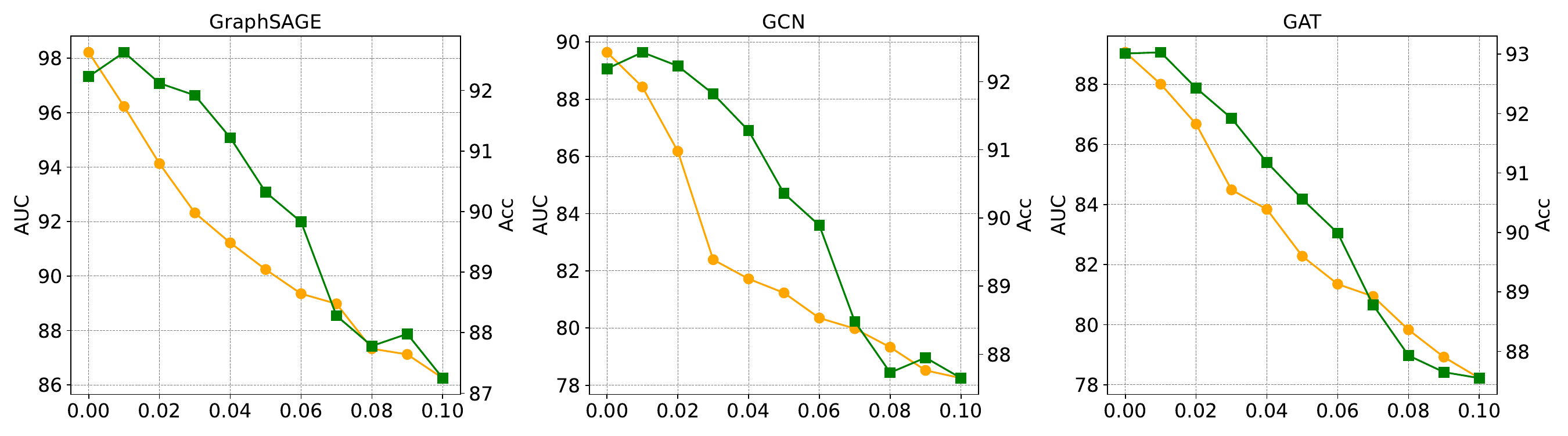}
    } 
  
   \caption{The attack performance of DMan4Rec on Amazon Photo dataset under defense methods.}
    \label{defense}
\end{figure*}

\textbf{Implementation Details.} 
We incorporate Laplacian noise and Gaussian noise~\cite{dpgn} into the model for defensive purposes. In specific, we add zero-mean Laplacian noise to the model, i.e., ${F_\theta }(x) = {f_\theta }(x) + Laplace(0,\lambda )$, and $\lambda$ is the Laplacian noise strength. We set the range of $\lambda$ from 0 to 0.1 to maintain the stability of the main task performance, so that it does not decline sharply. As for Gaussian noise, we set the noise range from 0 to 0.1. We conducted experiments with various levels of Gaussian noise and found that the performance of the main task and the attack performance fluctuated within a normal range. This allows us to better observe the balance between the two. We perform experiments to validate the effectiveness of potential defenses against DMan4Rec, as well as their impact on the primary task, as shown in Fig.~\ref{defense}.

\textbf{Results and Analysis.}
Fig.\ref{defense} shows the attack performance of DMan4Rec and the main task under the defense methods. 
Experimental results show that the performance of graph reconstruction drops as perturbation increases, and the decrease range is within 12\%. However, it is worth noting that the performance of the main task of the model will also decrease accordingly. At the same time, adding random noise has great instability in most settings, which is related to the randomness of the noise. Therefore, it is difficult to simultaneously trade off the performance of the main task and the performance of privacy protection just by adding noise to the model.


\subsection{Parameter Sensitivity (RQ6)}

In this section, we focus on the parameter sensitivity of DMan4Rec to different FedRecs. Specifically, we explore the impact of various loss term coefficients, label smoothing hyperparameters, and GNN model layers on the attack effect.

\begin{figure}[htbp!]

	\centering
	\includegraphics[width=1\linewidth]{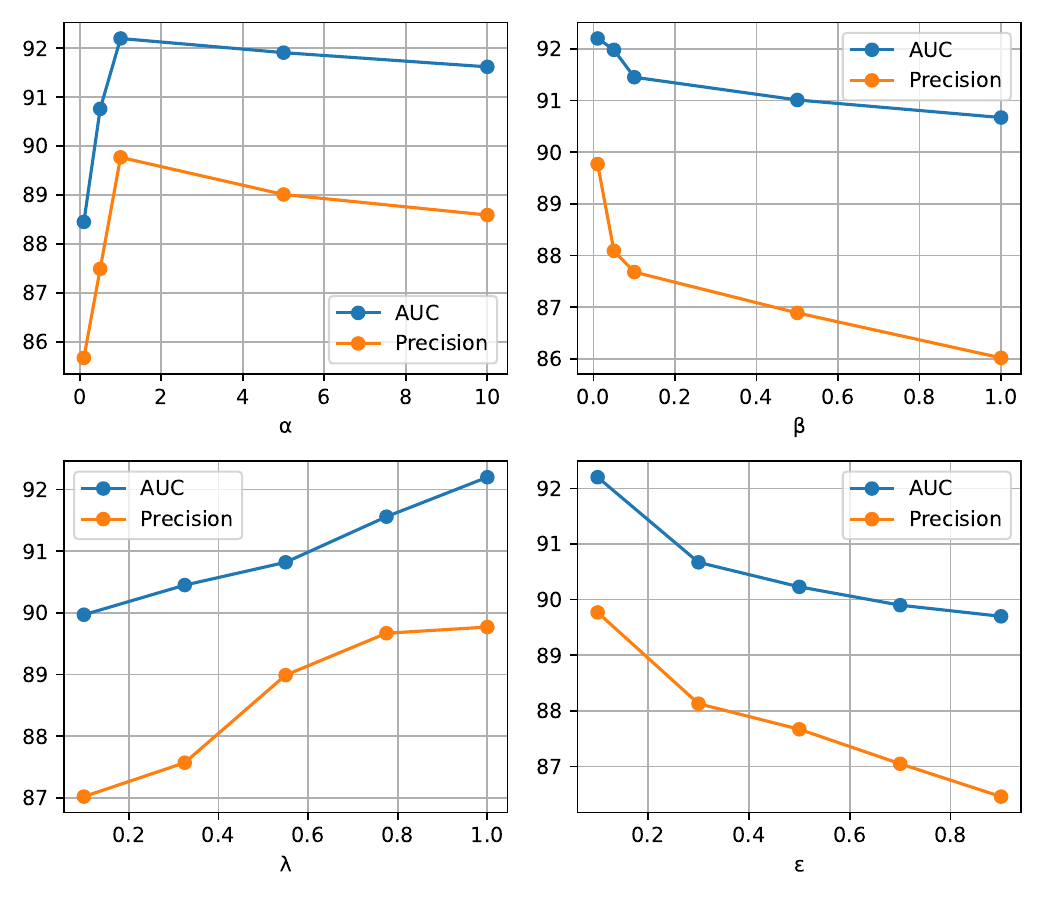}\\
	\caption{Comparison of DMan4Rec attack performance with different hyperparameters.}
	\label{fig6}
\vspace{-2em}
\end{figure}

\textbf{Implementation Details.} 
To validate the effect of various loss term coefficients and label smoothing hyperparameters on GRA performance, we conducted DMan4Rec with $\alpha $, $\beta$, $\lambda $ and $\varepsilon$ on Amazon computer dataset using GraphSAGE model. We examine the value of $\alpha $ in $\{0.1, 0.5, 1, 5, 10\}$, $\beta$ in $\{0.01, 0.05, 0.1, 0.5, 1\}$, $\lambda $ in $\{0.1, 0.325, 0.55, 0.775, 1.0\}$, and $\varepsilon$ in $\{0.1, 0.3, 0.5, 0.7, 0.9\}$, as shown in Fig.~\ref{fig6}, we present the results by measuring the AUC scores and Precision scores. 

\begin{figure}[htbp!]

	\centering
	\includegraphics[width=1\linewidth]{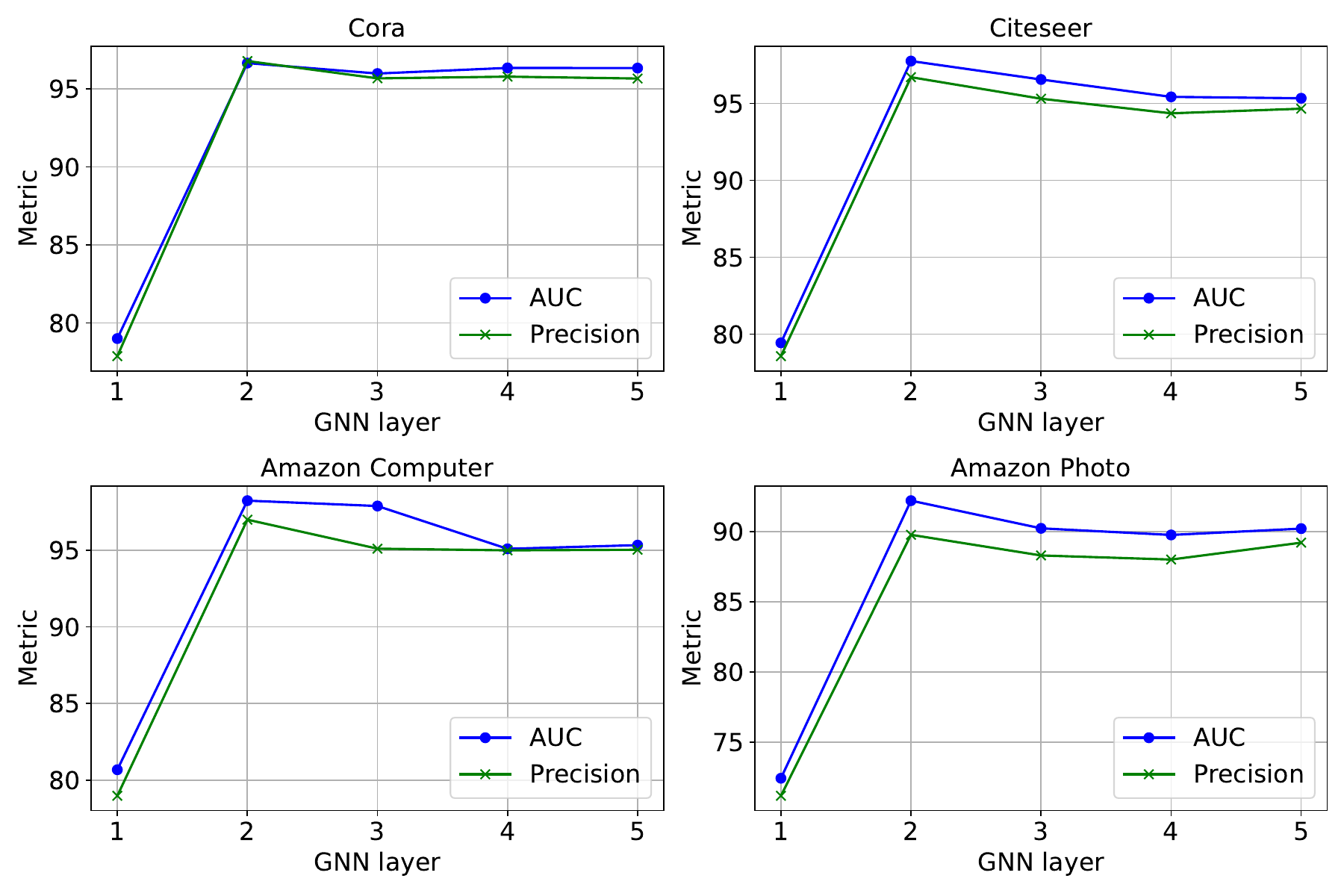}\\
	\caption{Comparison of DMan4Rec attack performance when the target model has different number of GNN layers.}
	\label{last}
\vspace{-2em}
\end{figure}

In addition, we conducted DMan4Rec on the target GNN model with 1, 2, 3, 4, and 5 GraphSAGE layers to validate whether the structure of the target model significantly affects the attack performance. We report the AUC and Precision scores on four datasets in Fig.~\ref{last}.

\textbf{Results and Analysis.}
Fig.~\ref{fig6} illustrates the impact of various loss coefficients and label smoothing hyperparameters on the attack effect in experiments conducted using the GraphSAGE model on the Amazon computer dataset. When formulating the target loss function, it is essential to consider the balance between attraction loss, repulsion loss, and cross-entropy loss by adjusting the corresponding regularization terms $\alpha$, $\beta$, $\lambda$, and $\varepsilon$. In the initial three figures of Fig.~\ref{fig6}, we compared the attack performance with different regularization weights. The optimal selection was found to be $\{ \alpha ,\beta ,\lambda ,\varepsilon \} = \{ 1,0.01,1,0.1\}$, where the weight was notably much smaller than others.
This is due to the imbalance in the number of linked and unlinked node pairs results in a higher repulsion loss. This choice aims to balance the effects of repulsion loss and attraction loss. As shown in the last figure of Fig.~\ref{fig6}, we observed that with the gradual increase in the set label smoothing hyperparameter, attack effectiveness diminishes. This indicates that the model becomes more resilient to adversarial samples. Moreover, it suggests that as the label smoothing hyperparameter increases, the model prioritizes predicting the correct category while disregarding information from other categories.

\vspace{0.13cm}
As shown in Fig.~\ref{last}, as the number of GNN model layers increases, GNN gradually aggregates information from multi-hop neighborhoods. Since more neighbor hops are considered, DMan4Rec also shows good performance accordingly. When the GNN has only one layer, the attack will be more difficult due to the lack of aggregate information between link nodes. Since no significant change in attack performance was observed when the number of model layers was larger than one, 2 GraphSAGE layers were adopted in this work.

\section{Conclusion}

In this paper, we firstly illustrate the limitations of existing graph reconstruction attacks for GFL, i.e., attack effectiveness, scalability and stealthiness. In order to solve these problems, we propose the first data manipulation aided reconstruction attack on GFL to explore the vulnerability of privacy leakage amplified by the malicious client. 
Extensive experiments on four general graph datasets show that DMan4Rec achieves SOTA GRA performance without affecting the node classification task performance. Moreover, DMan4Rec can transfer to the black-box setting and the complete overlap of the distribution graphs supports the stealthiness of the attack.

Unfortunately, we only study data manipulation assisted privacy leakage in the HGFL scenario for the case where one client is malicious, and have not conducted work on the VGFL scenario. Therefore, it is necessary to conduct privacy leakage work in different scenarios and consider malicious multiple parties. Moreover, this work warns that training datasets may expose privacy, and we call for more follow-up work to build a strong GFL framework to deal with such privacy leakage attacks. 
In addition, it will be appealing to further investigate privacy protection against malicious clients or malicious multiple parties.

\bibliographystyle{IEEEtran}      
\bibliography{graref}  


 




\end{document}